\documentclass[aps,  reprint, showpacs, amsmath, amssymb, groupedaddress]{revtex4-1}

\usepackage{graphicx}
\usepackage{dcolumn}
\usepackage{bm}
\usepackage{color}
\usepackage{ulem}
\newcommand{\ua}{\uparrow}
\newcommand{\da}{\downarrow}
\newcommand{\dg}{\dagger}

\begin{document}

\title{
Charge-density wave induced by combined electron-electron and electron-phonon interactions in 1$T$-TiSe$_2$: A variational Monte Carlo study
}



\author{Hiroshi Watanabe$^{1}$}
\email{h-watanabe@riken.jp}
\author{Kazuhiro Seki$^{2}$}
\author{Seiji Yunoki$^{1,2,3}$}
\affiliation{
$^1$Computational Quantum Matter Research Team, RIKEN Center for Emergent Matter Science (CEMS), Wako, Saitama 351-0198, Japan\\
$^2$Computational Condensed Matter Physics Laboratory, RIKEN, Wako, Saitama 351-0198, Japan\\
$^3$Computational Materials Science Research Team, RIKEN Advanced Institute for Computational Science (AICS), Kobe, Hyogo 650-0047, Japan
}


\date{\today}

\begin{abstract}
To clarify the origin of a charge-density wave (CDW) phase in 1$T$-TiSe$_2$, we study the ground state property of a half-filled 
two-band Hubbard model in a triangular lattice including electron-phonon interaction.
By using the variational Monte Carlo method, the electronic and lattice degrees of freedom are both treated quantum mechanically 
on an equal footing beyond the mean-field approximation.
We find that the cooperation between Coulomb interaction and electron-phonon interaction is essential to induce the CDW phase.
We show that the ``pure'' exciton condensation without lattice distortion is difficult to realize under the poor nesting condition of the underlying Fermi surface. 
Furthermore, by systematically calculating the momentum resolved hybridization between the two bands,
we examine the character of electron-hole pairing from the viewpoint of BCS-BEC crossover within the CDW phase and find that the strong-coupling BEC-like pairing dominates. 
We therefore propose that the CDW phase observed in 1$T$-TiSe$_2$ originates from a BEC-like electron-hole pairing. 
\end{abstract}


\maketitle

\section{Introduction}
Charge-density wave (CDW) is widely observed in low-dimensional solids and has been extensively studied both experimentally and 
theoretically~\cite{Gruner, Johannes}. 
Transition metal dichalcogenides MX$_2$ (M=transition metal, X=S, Se, Te) are one of the typical CDW materials with a layered 
triangular lattice structure.
They show various CDW patterns, depending on the combination of M and X~\cite{Wilson1, DiSalvo, Rossnagel}, and quite often superconductivity (SC) is observed next to the CDW phase
by applying pressure~\cite{Sipos, Kusmartseva}, doping~\cite{LJLi, Liu}, or intercalation~\cite{Morosan}.
However, the origin of CDW and SC has not been fully understood and it has been still under debate in spite of the long and 
extensive studies so far.

Recently, 1$T$-TiSe$_2$, one of the old transition metal dichalcogenides, has again attracted much interests in the context of 
exciton condensation. 
This material is a semimetal or a semiconductor in room temperature~\cite{GLi, Rasch, GMonney} and shows a commensurate 
CDW transition with a $2\times2\times2$ superstructure below $T_c\sim200$K.
The Fermi surface (FS) partially remains even in the CDW phase due to the imperfect opening of the energy gap.
The origin of the CDW phase is still controversial and the usual nesting mechanism seems unlikely due to its poor FS nesting~\cite{Zunger}.
Alternatively, exciton condensation has been proposed as a possible mechanism for the CDW phase~\cite{Wilson2, Cercellier, Cazzaniga}. 
Indeed, the large spectral weight transfer between Ti 3$d$ and Se 4$p$ bands and the flat energy spectrum just below the Fermi 
energy have been observed~\cite{Cercellier}, strongly suggesting the possibility of exciton condensation.
On the other hand, another possible mechanism for the CDW transition is a band Jahn-Teller effect which results from electron-phonon interaction~\cite{Hughes, Suzuki, Calandra, Zhu1}.
A large lattice distortion of several percent observed below $T_c$~\cite{DiSalvo} indicates 
the strong coupling between electronic and lattice degrees of freedom~\cite{Weber}.
Very recently, it is proposed that both mechanisms work cooperatively for the CDW transition~\cite{Taraphder, vanWezel1, CMonney1, Zenker1, Koley}.

Exciton condensation is a quantum state expected in low carrier density systems such as a semimetal or a semiconductor and has been 
extensively studied since 1960s~\cite{Mott, Kozlov, Jerome, Halperin}.
The exciton is a bound pair of an electron and a hole in different bands mediated by the interband Coulomb interaction.
When the binding energy of an exciton exceeds the band gap, 
the system has an instability toward condensation of excitons, namely, a spontaneous hybridization between different bands.
Since the repulsive Coulomb interaction is attractive between an electron and a hole, the exciton condensation is expected to occur in principle 
without considering any additional ``glue'' of the electron-hole pair.
Although extensive efforts have been devoted for half a century and several candidates have been 
proposed~\cite{Cercellier, Bucher, Mizuno, Wakisaka, Seki1}, the generally accepted materials for the exciton condensation are still absent.
Therefore, any conclusive evidence for the exciton condensation in real materials is highly desired for further progress and 
1$T$-TiSe$_2$ would be one of the promising examples.

In the pioneering studies for exciton condensation~\cite{Kozlov, Jerome}, an isotropic band 
dispersion with perfect FS nesting is assumed, for which the excitonic instability is always present 
in a semimetallic case.
The extension to anisotropic band dispersions shows that the degree of FS nesting greatly affects 
the instability of exciton condensation~\cite{Zittartz}.
Moreover, the mean-field approximation generally overestimates the instability toward ordered states, 
including the exciton condensation. 
Therefore, for discussion of the exciton condensation in real materials, the realistic band dispersion 
and the appropriate method beyond the mean-field approximation are both required. 

In this paper, a half-filled two-band Hubbard model with electron-phonon interaction in a triangular lattice is studied to understand the origin 
of the CDW phase in 1$T$-TiSe$_2$. The ground state properties are calculated using 
the variational Monte Carlo (VMC) method for multiorbital systems~\cite{Watanabe1}. 
We find that the cooperation between Coulomb interaction and electron-phonon interaction is essential to induce the CDW phase.
The CDW phase is observed between the normal metal and band insulator phases with intermediate interband Coulomb interaction.
We show that the ``pure'' exciton condensation without lattice distortion is difficult to realize under the poor FS nesting condition in a triangular lattice.
Furthermore, we systematically calculate the momentum resolved hybridization between the two bands 
to show that the strong-coupling BEC-like pairing dominates in the CDW phase. 
Our results therefore suggest that the CDW phase observed in 1$T$-TiSe$_2$ originates from the strong-coupling BEC-like electron-hole pairing. 

The rest of this paper is organized as follows.
In Sec.~\ref{Model}, a two-band Hubbard model in a two-dimensional triangular lattice is introduced as a low energy effective model 
for 1$T$-TiSe$_2$. 
The detailed explanation of the VMC method and the variational wave functions are also given in Sec.~\ref{Model}.
The numerical results are then provided in Sec.~\ref{Result}. 
Finally, the implication of our results for 1$T$-TiSe$_2$ is discussed in Sec.~\ref{Discussion}, 
followed by the summary in Sec.~\ref{Summary}. 

\begin{figure}[t]
\begin{center}
\includegraphics[width=\hsize]{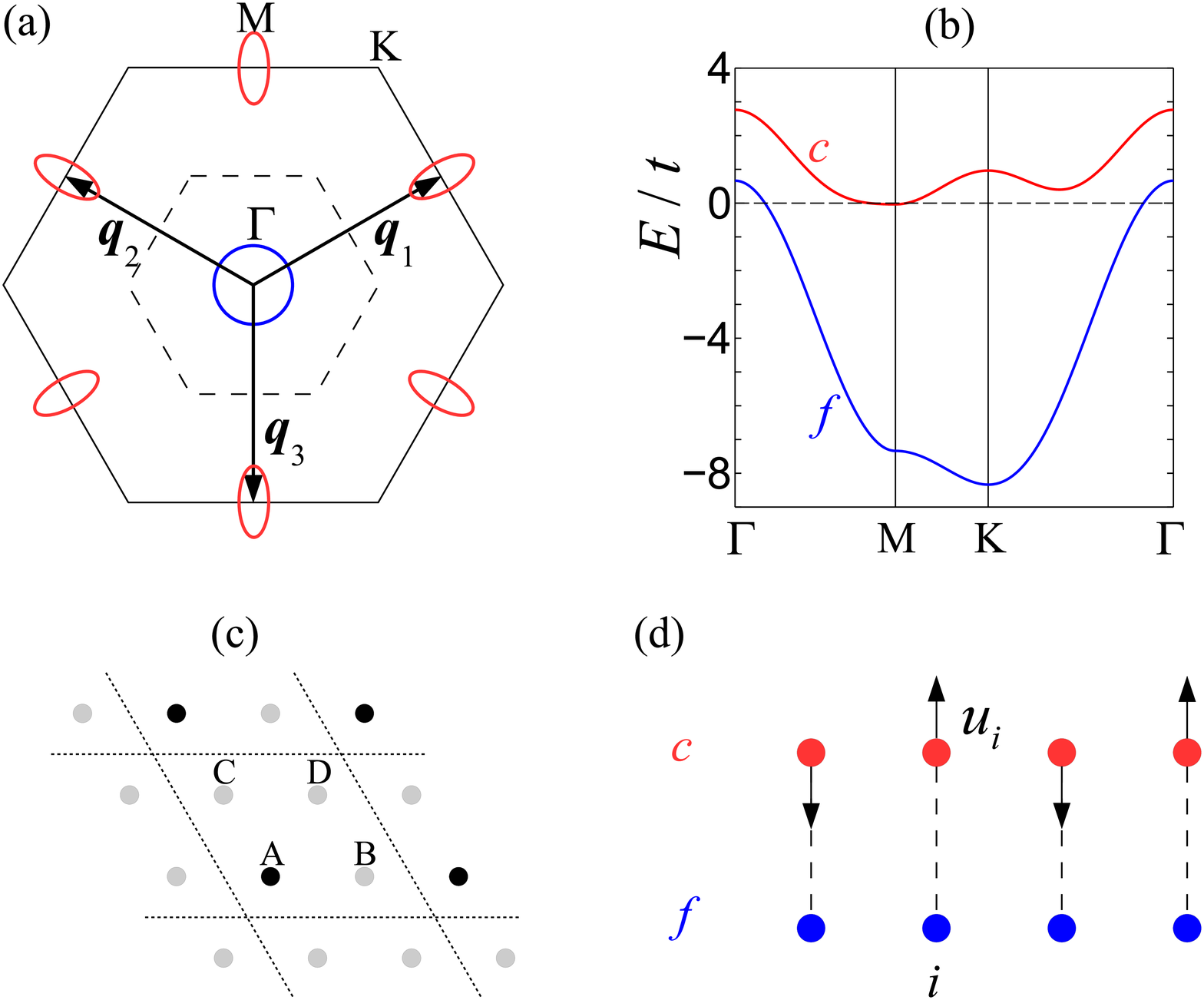}
\caption{(color online) 
(a) Fermi surfaces (blue circle and red ellipses) and (b) energy dispersions in the noninteracting limit for the two-band Hubbard model with 
electron density $n=2$. 
A set of tight-binding parameters used is $(t_f,\, t_c,\, t'_c,\, \mu_c) = (1.0,\,0.2,\,0.15,\,6.0)\,t$ with $t=t_f$ as an energy unit. 
(c) Schematic real-space figure of a 2$\times$2 superstructure (dotted lines) with each supercell containing 4 sites 
(A, B, C, and D). 
The corresponding ordering wave vectors $\bm{q}_1$, $\bm{q}_2$, and $\bm{q}_3$ are indicated in (a).
Dashed lines in (a) represent the folded Brillouin zone due to the formation of the 2$\times$2 superstructure in (c).
(d) Schematic real-space figure of the lattice distortion considered in the $c$-$f$ plane along 
one particular direction, e.g., AB direction indicated in (c). 
$u_i$ represents the displacement of $c$ atom at site $i$ from its original position. 
} \label{fig1}
\end{center}
\end{figure}

\section{Model and Method}\label{Model}
We consider a two-band Hubbard model in a two-dimensional triangular lattice defined as
\begin{align}
H&=\sum_{\bm{k},\sigma}\varepsilon^c_{\bm{k}}c^{\dg}_{\bm{k}\sigma}c_{\bm{k}\sigma} 
+\sum_{\bm{k},\sigma}\varepsilon^f_{\bm{k}}f^{\dg}_{\bm{k}\sigma}f_{\bm{k}\sigma} 
  \notag \\
&+U_{c}\sum_in^c_{i\ua}n^c_{i\da}+U_{f}\sum_in^f_{i\ua}n^f_{i\da}+U'\sum_in^c_in^f_i \notag \\
&+\frac{1}{\sqrt{N}}\sum_{\bm{k},\bm{q}, \sigma}
\left[g(\bm{k},\bm{q})(b_{\bm{q}}+b^{\dagger}_{-\bm{q}})
c^{\dagger}_{\bm{k}+\bm{q}\sigma}f_{\bm{k}\sigma}+\text{H.c.}\right] \notag \\
&+\sum_{\bm{q}}\omega(\bm{q})\left(b^{\dagger}_{\bm{q}}b_{\bm{q}}+\frac{1}{2}\right),
\end{align}
where $c^{\dg}_{\bm{k}\sigma}$ ($f^{\dg}_{\bm{k}\sigma}$) creates an electron in $c$ ($f$) band with momentum $\bm{k}$ and spin 
$\sigma\,(=\uparrow,\downarrow)$.
The band dispersions $\varepsilon^c_{\bm{k}}$ and $\varepsilon^f_{\bm{k}}$ are given as 
\begin{eqnarray}
\varepsilon^c_{\bm{k}}&=&2t_c\left(\cos k_x+2\cos \frac{1}{2}k_x \cos \frac{\sqrt{3}}{2}k_y\right) \nonumber \\
&+&2t'_c\left(\cos \sqrt{3}k_y+2\cos \frac{3}{2}k_x \cos \frac{\sqrt{3}}{2}k_y\right)+\mu_c
\end{eqnarray} 
and 
\begin{equation}
\varepsilon^f_{\bm{k}}=2t_f\left(\cos k_x+2\cos \frac{1}{2}k_x \cos \frac{\sqrt{3}}{2}k_y\right), 
\end{equation}
respectively.
We introduce the next-nearest-neighbor hopping $t'_c$ to locate the bottom of $c$ band at M points. 
$U_{c}$ $(U_{f})$ is an on-site intraband Coulomb interaction within $c$ ($f$) band and $U'$ is an on-site interband Coulomb interaction between $c$ and $f$ bands.
$n^{\alpha}_{i\sigma}$ is a number operator of $\alpha\, (=c, f)$ electron at site $i$ with spin $\sigma$ and 
$n^\alpha_i=n^\alpha_{i\ua}+n^\alpha_{i\da}$.
$g(\bm{k},\bm{q})$ is an electron-phonon coupling constant and $b^{\dg}_{\bm{q}}$ is a creation operator of phonon 
with momentum $\bm{q}$ and frequency $\omega(\bm{q})$.
In this model, the lattice distortion changes the $c$-$f$ bond length as shown in Fig.~\ref{fig1}(d),
which couples to the $c$-$f$ hybridization modulated 
with wave vector $\bm{q}$ through $g(\bm{k},\bm{q})$. 
The total number of sites is indicated by $N$.

The noninteracting tight-binding parameters, 
\begin{equation}
(t_f,\, t_c,\, t'_c,\, \mu_c) = (1.0,\,0.2,\,0.15\,,6.0)\,t, 
\end{equation}
are set to mimic the electronic structure of 
1$T$-TiSe$_2$ with a hole pocket at the $\Gamma$ point and electron pockets at the M points as shown in Figs.~\ref{fig1}(a) and \ref{fig1}(b).
The ordering wave vectors connecting $\Gamma$ and M points are denoted as $\bm{q}_1$, $\bm{q}_2$, and $\bm{q}_3$.
Although the ordering wave vectors observed in 1$T$-TiSe$_2$ connect $\Gamma$ and L points with a finite $k_z$ component~\cite{CMonney2},
here we consider a pure two-dimensional model for simplicity and the limitation of the model is discussed later.

The effect of Coulomb interaction and electron-phonon interaction is treated on an equal footing using a VMC method. 
We consider the trial wave function as follows:
\begin{equation}
\left|\Psi\right>=P_\text{e-ph}
\left|\Psi_{\text{ph}}\right>\left|\Psi_{\text{e}}\right>.
\end{equation}
$\left|\Psi_{\text{e}}\right>=P^{(2)}_\text{G}P_{\text{J}_{\text{c}}}\left|\Phi\right>$ is an electron wave function consisting of three parts. 
$\left|\Phi\right>$ is a Slater determinant constructed by diagonalizing 
the one-body part of Hamiltonian $H$ including the variational tight-binding parameters ($\tilde t_f=1$, $\tilde{t}_c, \tilde{t}'_c, \tilde{\mu}_c$) 
and the off-diagonal element $V$ which induces the $c$-$f$ hybridization.
Here, we assume $V=V_0\exp[-A(\tilde{\varepsilon}^c_{\bm{k}+\bm{q}_i}-\tilde{\varepsilon}^f_{\bm{k}})^2]$, and $V_0$ and $A$ are both 
variational parameters.
$V_0$ is an amplitude of the $c$-$f$ hybridization and $A$ controls the internal extent of exciton in $\bm{k}$ space.
We have found that the variational energy is improved by introducing $A$ and the behavior of $A$ is related to the BCS-BEC crossover of exciton condensation~\cite{Watanabe2}. 
$P^{(2)}_\text{G}$ is a Gutzwiller factor extended for two-band systems~\cite{Bunemann, Watanabe2}.
In $P^{(2)}_\text{G}$, possible 16 patterns of charge and spin configuration at each site $\left| \Gamma \right>$, i.e.,
$\left|0\right>=\left|0\;0\right>$, $\left|1\right>=\left|0\ua\right>$, $\cdots$, $\left|15\right>=\left|\ua\da\;\ua\da\right>$,
are differently weighted and their weight $\{g_{\Gamma}\}$ are optimized as variational parameters.
$P_{\text{J}_{\text{c}}}=\exp[-\sum_{i\neq j}\sum_{\alpha\beta}v^{\alpha\beta}_{ij}n^{\alpha}_in^{\beta}_j]$ is a charge Jastrow factor which controls long-range charge correlations. 
Here, $v^{\alpha\beta}_{ij}=v^{\alpha\beta}(|\bm{r}_i-\bm{r}_j|)$ is assumed and $\bm{r}_i$ is the position of site $i$. 

The trial wave function for phonon is assumed to be a Gaussian in the normal coordinate $\{Q_{\bm{q}}\}$ representation~\cite{Alder, Ohgoe},
\begin{equation}
\Psi^{\text{ph}}\equiv\bigl< \{Q_{\bm{q}}\} \bigl|\Psi^{\text{ph}}\bigr>=\exp\left[-\sum_{\bm{q}}\frac{1}{2}\frac{\left(Q_{\bm{q}}-\beta_{\bm{q}}\right)^2}{\alpha_{\bm{q}}^2}\right],
\end{equation}
where $Q_{\bm{q}}=\sum_iu_ie^{-i\bm{q}\cdot\bm{r}_i}/\sqrt{N}$ is Fourier transform of real space lattice distortion $\{u_i\}$ at site $i$. 
Since the ordering wave vectors $\bm{q}_1$, $\bm{q}_2$, and $\bm{q}_3$ are exactly half of the reciprocal lattice vectors of the normal phase 
[see Fig.~\ref{fig1}(a)], the corresponding normal coordinate $Q_{\bm{q}_i}$ ($i=1,2,3$) are real numbers.
Therefore, we can take the trial wave function and variational parameters $\alpha_{\bm{q}}$ and $\beta_{\bm{q}}$ all real. 
Notice that $\alpha_{\bm{q}}$ controls the extent of the Gaussian wave function, i.e., the amplitude of lattice vibration, and that 
$\beta_{\bm{q}}$ corresponds to the average value of $Q_{\bm{q}}$ and thus 
there exists a static lattice distortion with $\left< u_i \right>\neq 0$ for a finite $\beta_{\bm{q}}$. 
The Monte Carlo update scheme for $\{Q_{\bm{q}}\}$ and the estimation of phonon energy are the same as in Ref.~\onlinecite{Ohgoe}.

The remaining part is an electron-phonon projection operator: 
$P_{\text{e-ph}}=\exp\left[ \gamma\sum_iu_in^c_i(2-n^f_i)\right]$.
This operator controls the attraction between $c$ electrons and $f$ holes which results from the electron-phonon interaction and $\gamma$ is a variational parameter. 

The variational parameters in $\left|\Psi\right>$ are therefore 
$\tilde{t}_c$, $\tilde{t}'_c$, $\tilde{\mu}_c$, $V_0$, $A$, $\{g_{\Gamma}\}$, $\{v^{\alpha\beta}_{ij}\}$, 
$\{\alpha_{\bm{q}}\}$, $\{\beta_{\bm{q}}\}$, and $\gamma$, and they are simultaneously optimized using stochastic reconfiguration method~\cite{Sorella}.
The system sizes are varied from $L\times L=12\times12$ to $24\times24$ with antiperiodic boundary conditions in both directions of primitive lattice vectors for the triangular lattice.

\begin{figure}[t]
\begin{center}
\includegraphics[width=1.0\hsize]{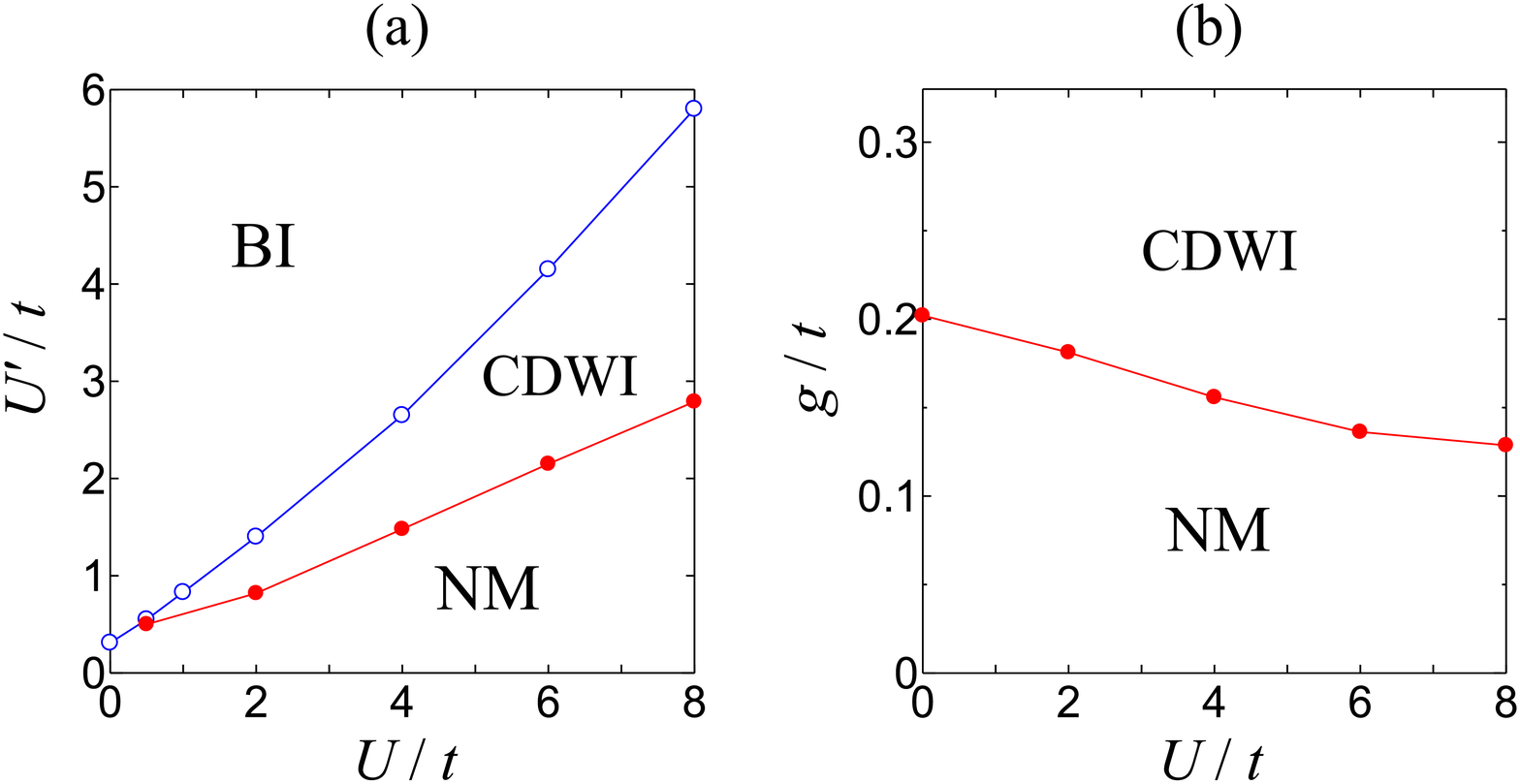}
\caption{\label{fig2}(color online)
Ground state phase diagram of the two-band Hubbard model in (a) $U$-$U'$ plane ($g/t=0.19$) and (b) $U$-$g$ plane ($U'=U/2$). 
We set $U_c=U_f=U$ and $\omega/t=0.1$. 
NM, CDWI, and BI denote normal metal, charge-density-wave insulator, and band insulator, respectively. The electron density is $n=2$, i.e., 
at half filling. 
} 
\end{center}
\end{figure}

\section{Result}\label{Result}
Figure~\ref{fig2}(a) shows the ground state phase diagram 
where $U_{c}=U_{f}=U$ and $U'$ are varied for fixed $g(\bm{k},\bm{q})/t=g/t=0.19$ and $\omega(\bm{q})/t=\omega/t=0.1$~\cite{note1}.
We find that there are three distinct phases in the phase diagram: normal metal (NM), charge-density-wave insulator (CDWI), and band insulator (BI).
When $U'$ is large enough, the $c$ band is lifted above the Fermi energy and the BI phase with the empty $c$ band and the fully-occupied 
$f$ band is stabilized. No static lattice distortion is observed in both NM and BI phases.

\begin{figure}[t!]
\begin{center}
\includegraphics[width=0.9\hsize]{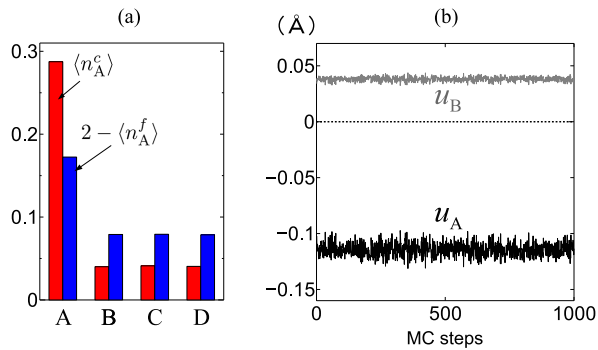}
\caption{\label{fig3}(color online)
(a) Average $c$-electron density $\left<n^c_{\text{X}} \right>$ (red bars) and $f$-hole density $2-\langle n^f_{\text{X}} \rangle$ (blue bars) 
in $2\times2$ unit cell for ${\rm X}={\rm A}$, $\rm B$, $\rm C$, and $\rm D$ [see Fig.~\ref{fig1}(c)]. 
(b) Snap shots of $c$-$f$ bond length $u_{\rm A}$ and $u_{\rm B}$ at A and B sites, respectively, as a function of Monte Carlo (MC) step. 
The model parameters used are ($U/t,U'/t,g/t,\omega/t$)=(4.0, 2.0, 0.19, 0.1) for $L=24$ in the CDWI phase. 
}
\end{center}
\end{figure}

Between the NM and BI phases, the CDWI phase emerges where the $c$-$f$ hybridization parameter
\begin{equation}
\Delta_{\bm{q}}=\sum_{\bm{k}, \sigma}\bigl<c^{\dg}_{\bm{k}+\bm{q}\sigma}f_{\bm{k}\sigma}+\text{H.c.}\bigr>
\end{equation}
is finite~\cite{note2} for $\bm q$ corresponding to the  three ordering wave vectors $\bm{q}_1$, $\bm{q}_2$, and $\bm{q}_3$, simultaneously, implying a triple-$\bm q$ CDW state.
Here, $\langle{\mathcal O}\rangle=\langle\Psi|{\mathcal O}|\Psi\rangle/\langle\Psi|\Psi\rangle$ with the optimized $|\Psi\rangle$.
Thus, the first Brillouin zone is folded as indicated in Fig.~\ref{fig1}(a). 
It leads to the charge disproportionation in 2$\times$2 unit cell [Fig.~\ref{fig1}(c)] with one charge rich A site and three charge poor B, C, and D sites [see Fig.~\ref{fig3}(a)].
In the CDWI phase, a static lattice distortion always occurs through the electron-phonon interaction 
and the ``pure'' exciton condensation without lattice distortion is never found.
Note that the CDWI phase is limited to a narrow region in Fig.~\ref{fig2}(a) especially for small $U/t$ in spite of a finite $g/t$. 
This is in sharp contrast with the case of a square lattice where the NM phase appears only at $U'=0$ and the exciton condensation phase is widely 
observed even without the electron-phonon interaction~\cite{Watanabe2, Zocher, Kaneko1}.
This difference is caused by the different FS nesting condition: the FS nesting is better (perfect if only with the nearest-neighbor hopping) 
in the square lattice but poor in the triangular lattice.
Therefore, the electron-phonon interaction is indispensable to manifest the CDWI phase under the poor FS nesting~\cite{Johannes}. 

We also show the phase diagram in Fig.~\ref{fig2}(b) where $U$ and $g$ are varied with $U'=U/2$.
The CDWI region is enlarged with increasing $U$ and $g$, implying that 
both Coulomb interaction and electron-phonon interaction stabilize the CDWI phase.
This result is thus qualitatively consistent with previous study~\cite{Zenker1}.
We also find that the CDWI phase is not stabilized, but only the NM and BI phases appear, when $g=0$, 
at least, in a realistic parameter region.
This suggests that the ``pure'' exciton condensation induced by the Coulomb interaction alone, the original idea of 
exciton condensation~\cite{Mott, Kozlov, Jerome}, is difficult to realize in our model.
The pure exciton condensation certainly occurs in particular models such as one-dimensional models~\cite{vanWezel1, Kaneko2} 
or two-dimensional models with perfectly nested electron and hole FSs~\cite{Watanabe2, Zocher, Kaneko1, Seki2}.
Therefore, the stability of the pure exciton condensation depends strongly on the lattice structure and the underlying FS.

Let us now examine the detailed properties of the CDWI phase.
Figure~\ref{fig3}(a) shows the distribution of average $c$-electron density $\left<n^c_{\text{X}} \right>$ and $f$-hole density 
$2-\langle n^f_{\text{X}} \rangle$ in 2$\times$2 unit cell (${\text{X}}$=A, B, C, and D).
It is found in Fig.~\ref{fig3}(a) that $c$ electrons and $f$ holes, i.e., mobile carriers, are concentrated mostly at A site 
with $\left<n^c_{\text{A}} \right> + \langle n^f_{\text{A}} \rangle >2$
to gain the $c$-$f$ hybridization energy coupled with the lattice distortion, 
while the number of these mobile carriers are small and $\left<n^c_{\text{X}} \right> +\langle n^f_{\text{X}} \rangle < 2$ 
at B, C, and D sites. 
Therefore, the system clearly exhibits the charge disproportionation. 
Notice that the mobile carrier densities at B, C, and D sites are the same within the statistical errors 
simply because the three ordering wave vectors $\bm{q}_1$, $\bm{q}_2$, and $\bm{q}_3$ 
are equivalent in a hexagonal lattice structure.

Next, let us discuss the lattice degrees of freedom in the CDWI phase. 
In the VMC calculation, the bond length $u_i$ always fluctuates around the average value during the Monte Carlo steps~\cite{Alder}. 
As shown in Fig.~\ref{fig3}(b), we find that the bond length at A site is shortened from the original one ($u_{\text{A}}<0$), 
while those at B, C, and D sites are lengthened (only $u_{\text{B}}>0$ is shown). 
Figs.~\ref{fig3}(a) and \ref{fig3}(b) thus confirm that the electronic and lattice degrees of freedom are strongly coupled and cooperatively induce 
the CDWI phase.

\begin{figure}[t!]
\begin{center}
\includegraphics[width=0.9\hsize]{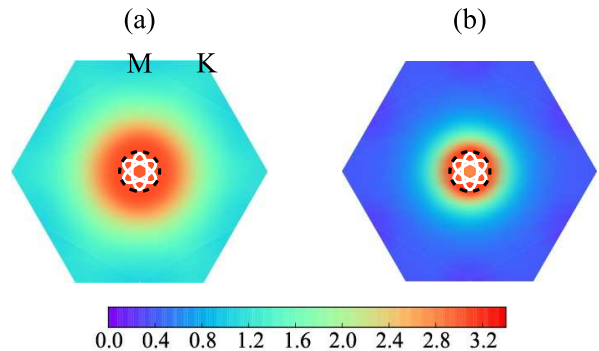}
\caption{\label{fig4}(color online) 
Momentum resolved $c$-$f$ hybridization $\phi(\bm{k})$ for (a) ($U/t,U'/t$)=(8.0, 4.0) and (b) ($U/t,U'/t$)=(3.0, 1.5).  
The noninteracting Fermi momentum $\bm{k}^c_{\text{F}}$ (folded around $\Gamma$ point at the center) and $\bm{k}^f_{\text{F}}$ are shown 
with white solid and black dashed curves, respectively [see also Fig.~\ref{fig1}(a)]. $g/t=0.19$ and $\omega/t=0.1$ are fixed for $L=24$. 
}
\end{center}
\end{figure}

Furthermore, we study the character of electron-hole pairing from the viewpoint of BCS-BEC crossover, which has been often discussed in 
the exciton problems~\cite{Nozieres, Bronold, Seki2, Zenker2, Kaneko1, Watanabe2}.
For this purpose, we calculate the momentum resolved $c$-$f$ hybridization $\phi(\bm{k})$ defined as 
\begin{equation}
\phi(\bm{k})=\sum_{\bm{q},\sigma}\bigl< c^{\dagger}_{\bm{k}+\bm{q}\sigma}f_{\bm{k}\sigma}+\mathrm{H.c.}\bigr>.
\end{equation}
Figure~\ref{fig4} shows $\phi(\bm{k})$ for ($U/t,U'/t$)=(8.0, 4.0) and (3.0, 1.5), 
both being located in the CDWI phase in Fig.~\ref{fig2}. 
For ($U/t,U'/t$)=(8.0, 4.0), $\phi(\bm{k})$ is extended in the whole Brillouin zone, indicating the strong-coupling BEC-like
pairing due to the large Coulomb interaction.
Although $\phi(\bm{k})$ becomes less extended with decreasing the Coulomb interaction, it still has a broad structure away from the 
Fermi momentum $\bm{k}_{\text{F}}$, as shown in Fig.~\ref{fig4}(b) for ($U/t,U'/t$)=(3.0, 1.5).
Indeed, the CDWI region rapidly decreases with decreasing $U$ [see Fig.~\ref{fig2}(a)] and our systematic calculations do not find 
a clear BCS-like region in the CDWI phase shown in Fig.~\ref{fig2}. 
Because of i) the poor nesting between $c$-electron and $f$-hole FSs and ii) the small density of states around the Fermi 
energy for low carrier densities, the energy gain due to the gap opening induced by the $c$-$f$ hybridization 
in the vicinity of $\bm{k}_{\text{F}}$ is small and hence the weak-coupling BCS-like pairing is not favored. 
Even in such a case, the BEC-like tightly-bounded electron-hole pairing in real space 
can be induced by the electron-phonon interaction with the help of Coulomb interaction and dominates the CDWI phase. 
In contrast, we have found a clear and wide BCS-like region for the same model but in a square lattice with perfectly nested FSs~\cite{Watanabe2}. 
Therefore, the FS nesting is essential for the BCS-like pairing.

\section{Discussion}\label{Discussion}
Finally, let us discuss the implication of our results for 1$T$-TiSe$_2$.
In our model, Ti 3$d$ and Se 4$p$ bands are simplified as $c$ and $f$ bands, respectively, and the orbital characters are ignored.
Moreover, our model only includes the change of $c$-$f$ bond length which couples with exciton condensation.
Even with these simplifications, our model captures the important energy scales of 1$T$-TiSe$_2$. 
The electron-phonon coupling used here is $4g^2/ \omega=0-3.6t\approx0-1.8$ eV, which is relevant for 1$T$-TiSe$_2$~\cite{Rossnagel} 
if we take $t\approx0.5$ eV.
The lattice distortion obtained in our calculation is $0.05-0.2${\AA}, consistent with the observed value $\sim 0.085${\AA}~\cite{DiSalvo, note3}.  
Our results thus suggest that the CDW phase observed in 1$T$-TiSe$_2$ is due to the strong-coupling BEC-like electron-hole pairing.
Indeed, the BEC-like character is indicated by several theoretical works~\cite{GMonney, Koley} and experimental observations 
such as a short coherence length estimated by Kohn anomaly~\cite{Rossnagel}, lack of incommensurate CDW phase,
relatively high electrical resistivity above $T_c$, and a large value of $2\Delta /k_{\text{B}}T_c$ 
($\Delta$: the CDW gap)~\cite{CMonney2}.

On the other hand, the chiral CDW phase observed in 1$T$-TiSe$_2$~\cite{Ishioka, Castellan} 
is beyond our model. In the chiral CDW phase, the charge density is modulated with clockwise or anticlockwise pattern. 
The proper description of this phase requires three dimensionality~\cite{vanWezel2} or higher-order electron-phonon and phonon-phonon
interactions~\cite{Zenker1} which induce the phase difference between the three ordering wave vectors. 
The origin of the SC induced by applying pressure or intercalation of Cu atoms is also an interesting unresolved issue.
The relation between the CDW and the SC is still controversial~\cite{Qian, Joe} and both conventional~\cite{Calandra, vanWezel1, SYLi} and unconventional SC~\cite{Ganesh} have been proposed.
Our results suggest that both electronic and lattice degrees of freedom are crucial to understand the origin of the SC.
Our study will be a first step toward the unified understanding of various quantum phases observed in 1$T$-TiSe$_2$.

\section{summary}\label{Summary}
In summary, we have studied the two-band Hubbard model in a triangular lattice for 1$T$-TiSe$_2$ with the electron-phonon interaction.
The VMC method is employed to treat the electronic and lattice degrees of freedom on an equal footing beyond the mean-field approximation.
We have shown that both Coulomb and electron-phonon interactions stabilize the CDW phase.
We have found that the ``pure'' exciton condensation without the lattice distortion is difficult to realize and the electron-phonon interaction is essential for the CDW phase. 
The character of electron-hole pairing within the CDW phase has also been examined by calculating the momentum resolved $c$-$f$ hybridization.
We have shown that the strong-coupling BEC-like pairing dominates the CDW phase.
Under the poor FS nesting condition and with small density of states around Fermi energy, 
the energy gain due to the gap opening in the vicinity of $\bm{k}_{\text{F}}$ is small and 
hence the weak-coupling BCS-like pairing is not favored. 
Our results thus conclude that the CDW phase observed in 1$T$-TiSe$_2$ originates from the strong-coupling BEC-like electron-hole 
pairing due to the cooperative Coulomb and electron-phonon interactions.

\section*{acknowledgment}
The authors thank Y. Fuseya, T. Shirakawa, T. Kaneko, and K. Imura for useful discussions.
The computation has been done using the RIKEN Cluster of Clusters (RICC) facility and the facilities of the Supercomputer Center,
Institute for Solid State Physics, University of Tokyo. 
This work has been supported by JSPS KAKENHI Grant No. 26800198 and in part by RIKEN iTHES Project.


\begin{thebibliography}{99}
\bibitem{note1} The system size used is $16\times 16$. However, we have checked that the phase boundary is almost unchanged for $L>16$.
\bibitem{note3} In our model, only two bands related to the exciton condensation are considered and the resulting CDW phase is an insulator.
In 1$T$-TiSe$_2$, the system remains metallic even when the CDW transition occurs. This is due to the imperfect opening of the gap.
Therefore, the lattice distortion obtained in our model tends to be larger than the experimental one.
\end{thebibliography}
\end{document}